# Nonlocal Heat Transport and Improved Target Design for X-ray Heating Studies at X-ray Free Electron Lasers

Oliver Hoidn and Gerald T. Seidler[(*)]

Physics Department, University of Washington, Seattle WA

The extremely high power densities and short durations of single pulses of x-ray free electron lasers (XFELs) have opened new opportunities in atomic physics, where complex excitation-relaxation chains allow for high ionization states in atomic and molecular systems, and in dense plasma physics, where XFEL heating of solid-density targets can create unique dense states of matter having temperatures on the order of the Fermi energy. We focus here on the latter phenomena, with special emphasis on the problem of optimum target design to achieve high x-ray heating into the warm dense matter (WDM) state. We report fully three-dimensional simulations of the incident x-ray pulse and the resulting multielectron relaxation cascade to model the spatial energy density deposition in multicomponent targets, with particular focus on the effects of nonlocal heat transport due to the motion of high energy photoelectrons and Auger electrons. We find that nanoscale high-Z/low-Z multicomponent targets can give much improved energy density deposition in lower-Z materials, with enhancements reaching a factor of 100. This has three important benefits. First, it greatly enlarges the thermodynamic parameter space in XFEL x-ray heating studies of lower-Z materials. Second, it allows the use of higher probe photon energies, enabling higher-information content X-ray diffraction (XRD) measurements such as in two-color XFEL operations. Third, while this is merely one step toward optimization of x-ray heating target design, the demonstration of the importance of nonlocal heat transport establishes important common ground between XFEL-based x-ray heating studies and more traditional laser plasma methods.

(*) seidler@uw.edu

**I. Introduction**

Dense matter under extreme conditions of pressure (P), temperature (T), or both, is a topic of classic and growing interest across multiple subfields of contemporary science. [1-5] We focus here on the very specific case of femtosecond-scale x-ray heating of crystalline matter, in which there is growing evidence that the lattice often has limited opportunity to structurally relax during the incident x-ray pulses [6-8] and that the loss of crystallinity during the x-ray pulse may have only modest scientific impact. [9] Such studies hold a significant and, we propose, unique position for discovery, because they encompass the case in which the consequences that traditional condensed phase electronic structure theory has on the structure of partially-ionized plasmas will be strongest and most easily interrogated. Hence, the study of crystalline matter at ambient density but highly elevated electronic temperature holds high potential for directly testing foundational issues in finite-T density functional theory, especially including the proper treatment of T-dependent functionals. [10-12]

This point has recently been made by Valenza and Seidler [12], who demonstrated that finite-T DFT makes strong, initially counter-intuitive predictions about the evolution of the absolute and relative Bragg peak intensities in x-ray diffraction (XRD) from crystalline matter as a function of electronic temperature on the $1 - 50$ eV scale. The key point is that XRD provides a more detailed interrogation of the population of electronic states for crystalline matter than it does for the more amorphous states interrogated after, e.g., laser shock heating. Furthermore, it is this temperature dependence that is a key *microscopic observable* of all finite-T DFT approaches: the central quantity calculated in DFT is, after all, the spatial distribution of electron density. Therefore, careful characterization of the real-space charge density at elevated electronic temperatures in a cool lattice gives a direct path to evaluating different DFT implementations. This is particularly significant as regards the temperature-dependent exchange functional, which is essential to predictions of bulk thermodynamic and elastic properties [10,11,13].

However, in such a research program there is a confounding detail. The most effective heating by x-rays will occur with lower-energy photons (that are more strongly absorbed) whereas any detailed interrogation of the real-space charge distribution by XRD requires the use of higher energy x-rays to obtain information over a wide momentum transfer range. [12] This dilemma raises a question that is new in the XFEL community but old in the broader plasma physics community: *Given the incident pulse characteristics and the desired sample material, how does one design a target to achieve optimal energy density deposition*?

The most comprehensive treatment of this question would include a fully spatio-temporal treatment of radiative transport as well as electronic dynamics and electron-atom interactions wherein, again because of the short time scales, lattice relaxation can be ignored or at least is secondary. Within this framework, the temporal evolution of electron-electron and electron-atom interaction includes several stages. First, the atomic physics of the core levels gives rise to an initial population of high-energy Auger electrons and photoelectrons that decay into low-energy ($<$ 50 eV) electronic excitations (both collective and single-particle) on the scale of a few femtoseconds. The resulting collective excitations decay by generating electron-hole pairs on the time scale of tens of femtoseconds. Subsequent electron-electron thermalization occurs on the scale of 100 fs – 1 ps for ambient matter [14-17], but in general has a strong eV-scale temperature dependence, thus requiring a self-consistent treatment at high incident flux levels.[15]

Here, we take a simpler approach with the goal of identifying and illustrating the most important contributors to x-ray heating and how their spatial extent strongly influences optimum x-ray heating target design, in the limited sense of optimizing the deposited net energy density in the desired sample phase. Specifically, we address the key questions surrounding nonlocal energy transport by hot electrons. This topic has a long history in plasma physics, especially for inertial confinement fusion target design, but enters here with typically lower-energy electrons, i.e., keV-scale, than are important in ICF and in direct-drive laser-heating studies. This causes the energy deposition length of the hot electrons to decrease from the 100-1000 μm scale for MeV electrons in laser experiments to instead only ~50-200 nm, depending on the atomic number of the species present in the XFEL x-ray heating target.

It is this much shorter length scale that brings us to consider multicomponent nanoscale targets for x-ray heating so that the influence of nonlocal energy transport by the hot electrons can be usefully engineered. While the importance of nanoscale energy transport has not previously been discussed in the context of XFEL heating target design, it has been studied and exploited in other experimental contexts. For example, there exists a significant body of literature in the medical physics community concerned with using gold nanoparticles for dose enhancement in radiotherapy treatment. [18,19] A contrasting application of nonlocal energy transport is found in the macromolecular crystallography community, where there is interest in the use of submicron incident x-ray beams so that a large fraction of high-energy electrons escape the beam spot before slowing down, thus reducing radiation damage in the probed sample volume. [20-24]

With the above context established, we consider here a nanostructured target design that enhances energy deposition in a sample material using nonlocal heat transport from a more strongly x-ray absorbing material in contact with the sample – we refer to this second material as a 'cladding' as a matter of convenience, for closer contact to the terminology of laser-shock target design, even when the geometry may not strictly be cladded. Fig. 1 sketches several corresponding geometries, but in the current paper we concentrate on the particularly simple one of Fig. 1 (c), consisting of a single thin film of sample material clad with Au. We use the Monte Carlo code PENELOPE to simulate three-dimensional electron-photon transport and the corresponding spatial distributions of deposited energy to demonstrate two benefits to the design: first, it significantly enhances in-sample energy deposition, and second, it relaxes constraints on XFEL pump photon energy in a way that substantially increases the information content of XRD measurements in certain experimental contexts.

We proceed as follows. In section II, we describe the methods used to simulate photoionization and electron transport in a nanostructured target and discuss the simplifying approximations on which we rely. In section III, we present and discuss simulation results of multilayer targets consisting of sample material clad on one or two sides with gold. We find that such a cladding configuration significantly increases deposited energy density in a sample material, with the largest enhancement in low-Z samples. We argue that this enhanced effect in low-Z samples opens the door to wide-angle x-ray diffraction (wide-angle XRD), with significant utility for studying the time dynamics of the energy relaxation cascade for both electronic and lattice/ion degrees of freedom in such materials. These observations are particularly relevant in the context of two-color x-ray pump x-ray probe experiments at XFELs[25-29], but also serve more generally to establish the importance of nanoscale nonlocal heat transport in high-intensity XFEL studies. Finally, in section IV we conclude.

## II. Methods

The simulation of electron transport in condensed matter is an area of ongoing research. In addition to continuing development of well-established codes in the high-energy experimental particle physics community [30,31], new developments include incorporation of *ab initio* band structure calculations in order to accurately model the electron mean free paths of interband transitions and plasmon excitations from relativistic energies down to a few eV. [32,33]

In the regime relevant to the present study, calculation of the spatial distribution of deposited energy caused by absorption of a hard x ray requires accurate treatment of the processes that describe scattering of photo- and Auger electrons at the 100 eV to 10 keV scale (generation of secondary x-ray photons, though present, plays a negligible role in energy transport). The simplest atomic treatments of elastic and inelastic scattering demonstrate that, for mid- and high-Z elements, the ratio of elastic to inelastic total cross sections is of order unity and that characteristic elastic scattering angles are sufficiently large (for instance, of order 1 rad for ~1 keV electrons) to influence deposited energy distributions. [34] Both components, therefore, must receive accurate treatments to adequately model spatial energy deposition distributions in a nanostructured target.

The spatial distribution of deposited energy is determined by the electron stopping power $dE/dz$, which in a classical treatment is related to a material's dielectric function $\varepsilon(q, \omega)$ by

$$\frac{dE}{dz} = \frac{2\hbar^2}{\pi a_0 m_0 v^2} \iint \frac{q_y \omega \, Im[-1/\varepsilon(q,\omega)]}{q_y^2 + (\omega/v)^2} dq_y d\omega, \quad (1)$$

where $\omega$ is angular frequency, $q$ is momentum transfer (with $q_y$ the magnitude of the component for momentum transfer perpendicular to the z-direction), $a_0$ is the Bohr radius, $m_0$ is the electron mass, and $v$ is the electron velocity. [34] In the case of electron showers generated by 5-10 keV photons, the electron stopping power's dependence on $v$ causes nonlocal energy transport to be dominated by the highest-energy Auger and photoelectrons. Though the slower time evolution of the subsequent electronic and lattice dynamics may be neglected in the present context of simulating fsec-scale energy transport, the possibility of interrogating it by time-resolved XFEL pump-probe measurement is an interesting topic in its own right.[25,26]

To model the above physics we used the code PENELOPE, which implements particle-tracking Monte Carlo simulations of electron showers generated by x-ray photoionization.[31] PENELOPE uses total and differential cross sections based on several physical models. Briefly, it derives elastic and inner-shell inelastic cross sections from strictly atomic wave functions, while the valence contribution to the inelastic double differential cross section is based on the Born approximation and generalized oscillator strength model of Liljequist [35,36], with an energy loss-dependent normalization that allows the model to replicate empirical stopping power data (provided as program input). Although the inelastic scattering cross section is dominated by low-energy loss collisions, inner shells contribute the majority of the stopping power for several-keV electrons, which account for the longest-range energy transport. For electrons of those energies the stopping power of a compound may be approximated within five percent by a stoichiometric sum based on atomic treatments of its constituents (an observation referred to as Bragg's rule). [37] Consequently we employed material data files generated by the PENELOPE 2011 program MATERIAL, which applies this approximation to infer stopping powers of arbitrary compounds using data from the NIST ESTAR database. [31,38]

**III. Results and discussion**

We now present results for several realizations of our nanostructured target design, all of which consist of thin films clad with Au on one or both sides. The heating of an Fe thin film via nonlocal heat transport by hot electrons is illustrated in Fig. 2, which shows a two-dimensional projection of electron trajectory traces in an Au-Fe-Au trilayer stimulated with 7 keV incident photons. The color-coding of the tracks shows that, due to the much larger number of photoexcitations in the Au cladding compared to Fe inclusion, most hot electrons propagating in the Fe are part of a photoionization relaxation cascade originating in the cladding. Inelastic scattering of these hot electrons is the dominant contribution to energy deposition in the central Fe region, as quantified by Fig. 3, which compares the linear energy deposition of several Au-Fe-Au trilayer configurations to that in bare Fe.

Photoionization by 7 keV photons yields mean energy deposition lengths $l$ of 15.0 nm and 35.3 nm, respectively, in simulated bare Fe and bare Au targets, where $l = \int_{z=0}^{\infty} z(\vec{r})\, \rho(\vec{r})\, d^3\vec{r}$, with $\rho(\vec{r})$ the volume density of deposited energy and $z$ the magnitude of the projection of $\vec{r}$ onto a fixed, arbitrary unit vector. Consistent with the above characteristic lengths, we found that absorbed energy density in the Fe inclusion saturates beyond an Au cladding thickness of 50 nm. Fig. 3 (a) shows the deposited energy distribution in a bare $Fe_3O_4$ target and in several Au-$Fe_3O_4$-Au trilayers with varying thicknesses of the $Fe_3O_4$ inclusion. An interior layer thickness of 50 nm results in a factor of five enhancement in deposited energy density relative to the bare $Fe_3O_4$ target. The increase in deposited energy density in a clad sample compared to a bare one is significantly larger for lower-Z materials, reaching a factor of 100 for an Au-C-Au target of the same geometry (Fig. 4).

These enhancements in energy deposition increase the accessible thermodynamic parameter space in all XFEL heating experiments, which is particularly significant for experimental diagnostics that require deviation from optimal pump pulse characteristics and are therefore normally incompatible with heating studies, for instance XRD. We illustrate this in Fig. 5, which compares the energy deposition in Au-Fe-Au and Au-$Fe_3O_4$-Au targets stimulated with photons below the *K*-edge of Fe to that in a bare Fe target heated by photons above the edge. Nonlocal heating of the former samples compensates for the reduction in sample heating caused by lowering the incident photon energy below the Fe *K*-edge; the multicomponent targets thus allow improving the ratio of signal to (fluorescence) background while—in the more favorable case of $Fe_3O_4$–maintaining an energy deposition density comparable to the highest level possible with an equivalent monolithic target. However, Fig. 6 also demonstrates a tradeoff of the cladding's presence: the diffracted signal from Au is stronger than that from the sample, making the described reduction in background worthwhile only assuming sufficient separation between Bragg peaks of the sample and cladding.

Low-Z sample materials provide a separate, independently interesting, case for the use of structured target design in XRD studies. In such materials, nonlocal heat transport is effective over a much wider range of incident photon energies compared to direct x-ray absorption. Until now, x-ray heating studies of low-Z materials, such as graphite, have required incident photon energies below 3 keV to reach HED conditions (> ~1 eV temperatures) due to these materials' small photoelectric cross sections in the hard X ray (photon energy > 5 keV) regime. This restriction limits the kinematically accessible range of momentum transfers in XRD, which correspondingly reduces available information on real-space charge density.

This creates an experimental dilemma with scientific consequences. For example, Hau-Riege et al.[39] showed evidence for ultrafast melting of graphite during a 40 fs-long XFEL pulse but were limited, for the reason described above, to using 2 keV incident photons, yielding diffraction from only the 002 Bragg reflection of graphite. The authors interpreted quenching upon heating of the 002 peak as evidence of nonthermal lattice melting. However, Valenza *et al.*[12] questioned this conclusion based on simulated diffraction using frozen-core finite-T DFT calculations, which predicted strong quenching of the graphite 002 reflection due to purely electronic reorganization in crystalline graphite at 10 eV electronic temperature. In graphite and other low-Z systems, the only means of unambiguously separating lattice disorder from electronic heating in the XRD signal is to probe several Bragg peaks, including the lowest-order reflections and their harmonics. [12]

It is therefore interesting to ask whether high energy-densities can be achieved in graphite when using photons suitable for wide-angle scattering. In Fig. 4 we show the deposited energy densities in Au-C-Au trilayers of several interior thicknesses, once again using 7 keV incident photons (sufficient to probe the 006 reflection of graphite). The deposited energy density in the interior layer is at least a factor of 100 greater compared to an unclad sample with the same incident photon energy, and a factor of two greater compared to an unclad sample stimulated with 2 keV photons. Indirect heating via high-Z cladding thus eliminates the constraint of selecting incident photon energies near a low-Z material's small core binding energies, making wide-angle XRD possible. In the context of carbon, the weakness of the XRD from C compared to Au can be uniquely compensated with a highly-oriented pyrolytic graphite (HOPG) sample, whose high-reflectivity 00*l* peaks yield much higher signal to background ratios than the powder-like Bragg and thermal diffuse scattering of polycrystalline Au. Similar configurations exploiting mosaic or single-crystal samples may enhance wide-angle XRD on a variety of low-Z systems, offering a much-improved ability to experimentally test predictions of finite-T DFT-based modeling of electronic structure in low-Z condensed matter, where finite-T effects are easiest to identify because of the relatively large valence-electron contribution to the XRD signal.[12]

The simulations presented in this paper constitute a first demonstration of a particularly simple implementation of structured target design. One can imagine several improved designs that achieve the same level of nonlocal sample heating while averting some of the disadvantages of our multilayer approach. For example, a uniform mixture of small (< 50 nm diameter) sample and heater nanoparticles would show similar mean deposited energy densities to a multilayer target and can be prepared by, e.g., spin coating or drop-casting. Such targets would have more homogeneous heating and would additionally allow preparation of much thicker targets and give much higher scattered intensities. A similar result may be possible using electrochemical or vapor deposition to embed sample materials inside porous high-Z metal substrates.[40,41] Two-color XFEL experiments may also lend themselves to lithographically patterned designs with concentric cylindrical volumes of (inner) sample and (outer) cladding materials, wherein the more tightly-focused probe pulse would be inscribed in a volume free of cladding material. Such a configuration would have the intention of reducing (cladding) background relative to signal, which would be particularly useful for weakly-diffracting low-Z samples.

**IV. Conclusion**

We model the spatial distribution of deposited energy in nanostructured targets for hard x-ray XFEL heating experiments using the Monte Carlo code PENELOPE. We find that two-component targets consisting of a sample material and high-Z cladding achieve substantial nonlocal heating of the sample via the relaxation cascade following transport of multi-keV Auger and photoelectrons. We argue that this target design approach will bring substantial benefits to XFEL heating experiments in the following ways: first, by enlarging their accessible thermodynamic parameter space and second, by improving the capability of x-ray diffraction diagnostics to characterize finite-temperature electronic structure and to distinguish between thermalization of the electronic and lattice degrees of freedom in crystalline warm dense matter systems.

**Acknowledgements**

We thank Joshua Kas for useful discussions. This work was supported by the United States Department of Energy, Basic Energy Sciences, under grant DE-SC00008580 and by the Joint Plasma Physics Program of the National Science Foundation and the Department of Energy under grant DE-SC0016251.

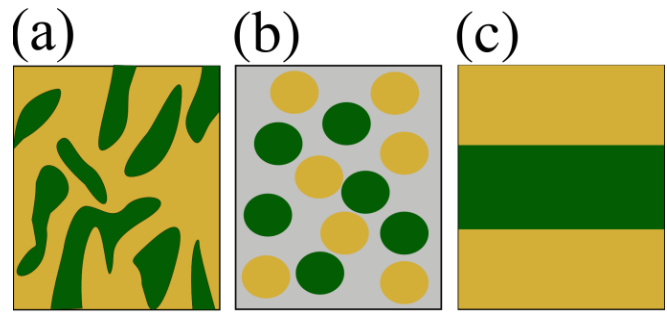

**Fig. 1**. Representations of three types of multicomponent targets composed of sample material (green) and heater cladding (yellow). (a): A porous subtrate filled with sample material; (b): a mixture of cladding and sample nanoparticles embedded in a solid matrix; (c): a multilayer film.

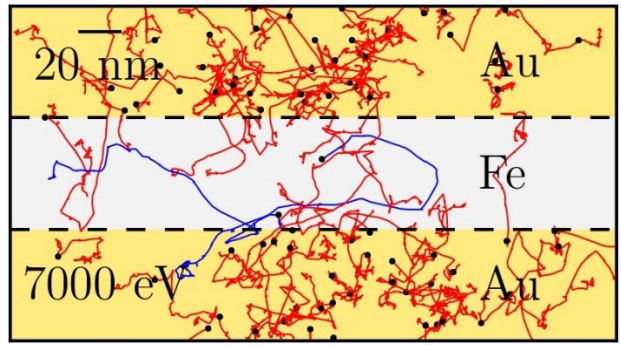

**Fig. 2**. Visualization of a 3-D Monte Carlo simulation of electron transport in an Au-Fe-Au target heated by 7 keV photons, incident normally from the top of the page. Electron tracks are projected onto the plane of the page; showers resulting from photoexcitation of Au and Fe atoms are red and blue, respectively. Note that most of the electron tracks in the Fe are due to absorption events in the Au.

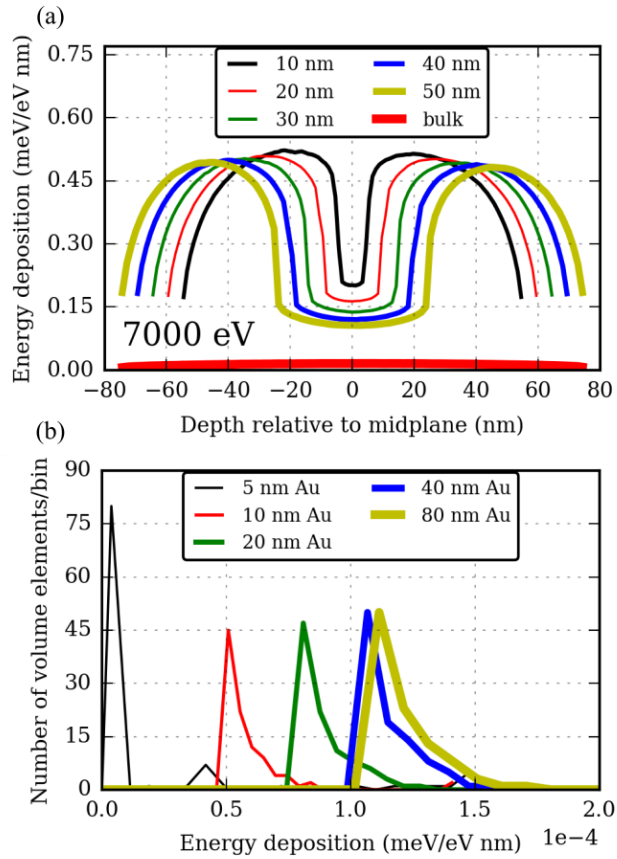

**Fig. 3**. (a) Linear energy deposition density generated by 7 keV photons incident on an Au-$Fe_3O_4$-Au target, displayed for several thicknesses of the central $Fe_3O_4$ layer and a fixed Au cladding thickness of 50nm. (b) Histograms of energy deposition density in volume elements of the $Fe_3O_4$ inclusions in Au-$Fe_3O_4$-Au targets, displayed for several thicknesses of the Au cladding and a fixed $Fe_3O_4$ layer thickness of 50 nm.

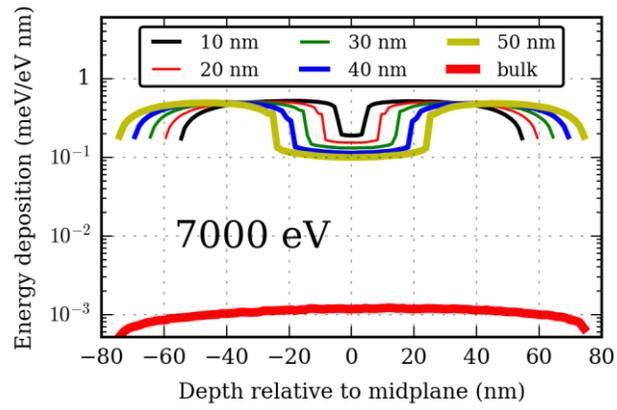

**Fig. 4**. Linear energy deposition due to 7 keV photons incident on an Au-C-Au target displayed for several thicknesses of the central C layer and a fixed outer cladding thickness of 50nm.

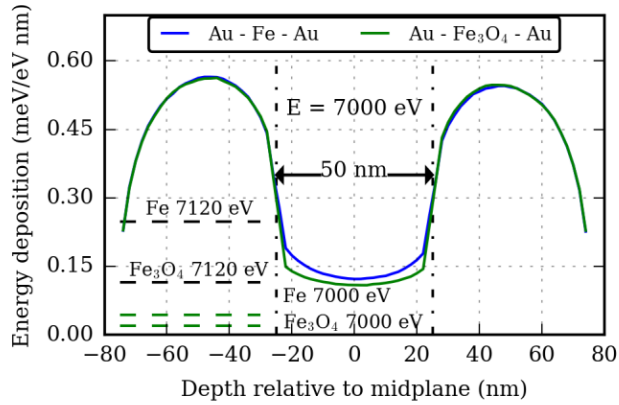

**Fig. 5.** Linear energy deposition in layered Au-Fe-Au and Au-$Fe_3O_4$-Au targets of 150 nm total thickness stimulated by 7 keV photons. Dashed lines indicate energy deposition in bulk $Fe_3O_4$ and Fe at photon energies of 7.12 keV (above the iron K-edge) and 7 keV (below the edge). The multilayer configuration sufficiently enhances energy deposition so as to partially compensate for the difference between pre- and above-edge x-ray photoelectric cross sections. The benefit is particularly pronounced in $Fe_3O_4$ due to its much lower density and photoelectric cross-section.

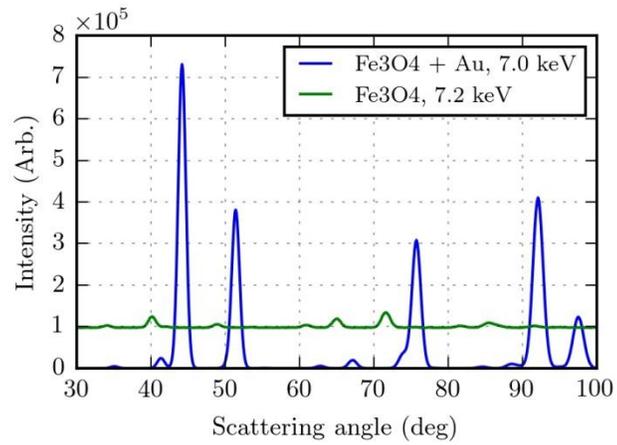

**Fig. 6**. Simulated powder diffraction of 50 nm Au-50 nm Fe$_3$O$_4$-50 nm Au stimulated by X-rays below the Fe K-edge (blue) compared to that resulting from photons above the edge incident on bare Fe$_3$O$_4$, including fluorescence background(green).